# The implications of collisions on the spatial profile of electric potential and the space-charge limited current

Allen L. Garner[1,2,3,a)] and N. R. Sree Harsha[1]

[1] School of Nuclear Engineering, Purdue University, West Lafayette, Indiana 47907, USA

[2] Department of Agricultural and Biological Engineering, Purdue University, West Lafayette, Indiana 47907, USA

[34] Elmore Family School of Electrical and Computer Engineering, Purdue University, West Lafayette, Indiana 47907 USA

[a)] **Author to whom correspondence should be addressed:** algarner@purdue.edu

The space-charge limited current (SCLC) in a vacuum diode is given by the Child-Langmuir law (CLL), whose electric potential $\phi(x) \propto (x/D)^{4/3}$, where $x$ is the spatial coordinate across the gap and $D$ is the gap separation distance. For a collisional diode, SCLC is given by the Mott-Gurney law (MGL) and $\phi(x) \propto (x/D)^{3/2}$. This Letter applies a capacitance argument for SCLC and uses the transit time from a recent exact solution for collisional SCLC to show that $\phi(x) \propto (x/D)^{\xi}$ for a general collisional gap, where $4/3 \leq \xi \leq 3/2$. Furthermore, $\xi$ is strictly a function of $\nu T$, where $\nu$ is the collision frequency and $T$ is the electron transit time. Using this definition of $\xi$, we estimate the spatial dependence of the electron velocity and use the capacitance to derive an analytic equation for collisional SCLC that agrees within ~5-6% of the exact solution that requires solving parametrically through $T$. We derive equations in the limits of $\nu \to 0$ and $\nu \to \infty$ for general $\xi$ that asymptotically recover the CLL as $\nu \to 0$ and the MGL as $\nu \to \infty$. Matching these limits shows that $\xi \approx 1.40$ and $V \propto D^2 \nu^2$ at the transition from a vacuum to collisional diode for any device condition.

Characterizing electron emission is critical in many applications, including high-power microwaves, directed energy, nano vacuum transistors, and time-resolved electron microscopy [1]. Electrons may be emitted in numerous ways [2-4], including field emission [2-6], thermal emission [7-9], and photoemission [9,10]. Regardless of the source, the maximum current permissible in a diode is given by the space-charge-limited current (SCLC) [1,11]. For emission from an infinitesimally small patch on an infinitely large planar electrode [12, 13], the Child-Langmuir law (CLL) is given by [1]

$$J_{CL} = \frac{4\sqrt{2}}{9}\epsilon_0 \sqrt{\frac{e}{m}} \frac{V^{3/2}}{D^2},\qquad(1)$$

where $\epsilon_0$ is the permittivity of free space, $V$ is the electric potential drop between the cathode and anode, $D$ is the gap distance, $e$ is electron charge, and $m$ is the electron mass. This has been extended to more realistic geometries [14] to derive the one-dimensional (1D) SCLC for concentric cylinders [15-17], concentric spheres [15,17], tips [17,18], and curvilinear geometries [16]. Further studies modeled multiple dimensions, ranging from emission from a finite patch on a longer electrode [19, 20] to emission from a full electrode in two- and three-dimensions by accounting for the resulting fringing fields through vacuum capacitance [21]. The importance of thermal emission in diodes has motivated the extension of the CLL to nonzero initial velocities [17, 22-28], demonstrating the existence of a bifurcation solution characterized by electron reflection and a true SCLC representing the maximum current permissible in the gap [25]. Halpern *et al*. applied variational calculus to extend the bifurcation SCLC solutions to nonplanar diodes [26], while Harsha *et al*. used Lie point symmetries to derive the SCLC for nonzero initial velocity for multiple geometries [17].

While valuable, these analyses focused on vacuum. More recent studies have explored breakdown at microscale and smaller gap distances at atmospheric pressure for numerous applications, including medicine, sterilization, and combustion [3,4,29-32]. The strong electric field at breakdown under these conditions causes electron field emission from the cathode [3, 4, 30-32]. These electrons ionize the gas near the cathode. The resulting ions create a space-charge field that enhances field emission; they also strike the cathode to release more electrons to provide further feedback. This decreases the breakdown voltage, deviating from Paschen's law [4]. Reducing the gap distance below 1 $\mu$m at atmospheric pressure approaches the transition between the Fowler-Nordheim equation (FNE) for field emission, the CLL, and the Mott-Gurney law (MGL) for collisional SCLC [3,6], given by

$$J_{MG} = \frac{9}{8}\mu_e \varepsilon_0 \frac{V^2}{D^3}, \qquad (2)$$

where $\mu_e$ is electron mobility [33]. Experiments demonstrated the transition to SCLC for submicroscale gaps at atmospheric pressure [34] and further theoretical studies incorporated quantum effects below 100 nm [35]. Many studies fit (2) to measured $J - V$ curves to determine $\mu_e$ in semiconductors [36-41], which further motivates understanding the implications of collisions on the SCLC.

Microscale breakdown and the potential loss of vacuum in high-power devices due to contaminants or leakage motivate further characterization of collisional SCLC. Numerous studies have explored the theoretical transitions between electron emission mechanisms and SCLC [3, 5, 7, 8, 9, 10, 23, 42, 43]. Because incorporating temperature requires accounting for nonzero initial velocity for the SCLC, recent studies examined this limit with the MGL by linking field emission and thermal emission directly [8] and separately by deriving the MGL with nonzero initial velocity by using the electron transit time [46]. The latter approach [46] derived an exact solution for the SCLC from pure vacuum to infinitely collisional with nonzero initial velocity.

While the SCLC can be assessed over a wide range of parameters by using this exact solution [46], it does not provide intuition into the functional dependences because the voltage and current are defined parametrically through the transit time $T$. This Letter derives a simpler relationship to link SCLC from the CLL to the MGL by leveraging vacuum capacitance, which was used to individually recover the CLL [47] and MGL [48]. We obtain the spatial profile of the electric potential as a function of the number of collisions and information concerning the transition from vacuum to collision dominated SCLC that was not obvious from the exact solution, while providing a rapid means to calculate the SCLC.

We consider a 1D planar diode filled with a neutral gas with constant collision frequency $\nu$. The diode contains a grounded cathode with electric potential $\phi = 0$ at $x = 0$ and an anode at $x = D$ held at an electric potential $\phi = V$. An electron is emitted from the cathode with velocity $u(0) = u_0 = 0$ and initial acceleration $u'(0) = eE_s/m$, where $E_s$ is the electric field at the cathode. Poisson's equation is given by

$$\frac{d^2\phi}{dx^2} = \frac{\rho}{\epsilon_0}, \tag{3}$$

where $\rho$ is the charge density and $\epsilon_0$ is the permittivity of free space. The electron force law is given by

$$m\frac{du}{dt} = e\frac{d\phi}{dx} - m\nu u, \tag{4}$$

where the first term on the right-hand side represents the acceleration due to the electric field and the second term represents friction due to electron collisions with the neutral gas. To simplify the equations, we define the following nondimensional parameters to be consistent with prior studies [46] to facilitate comparisons:

$$\phi = \phi_*\bar{\phi}; \quad J = J_*\bar{J}; \quad x = x_*\bar{x}; \quad t = t_*\bar{t}; \quad E = E_*\bar{E}; \quad \nu = \nu_*\bar{\nu}; \quad \phi_* = E_*x_*; \quad J_* = A_{FN}B_{FN}^2;$$

$$x_* = \frac{e\epsilon_0^2}{mA_{FN}^2 B_{FN}}; \quad t_* = \frac{\epsilon_0}{A_{FN}B_{FN}}; \quad \nu_* = \frac{A_{FN}B_{FN}}{\epsilon_0}; \quad E_* = B_{FN}; \quad u_* = \frac{x_*}{t_*}, \tag{5}$$

where the bars represent dimensionless parameters, the terms with subscript * are scaling parameters, $A_{FN}$ and $B_{FN}$ are Fowler-Nordheim (FN) coefficients given by $A_{FN} = e^3/(16\pi^2\hbar\Phi)$ and $B_{FN} = (4\sqrt{2m\Phi^3})/(3\hbar e)$, $\Phi$ is the electrode work function, and $\hbar$ is the reduced Planck's constant. Table I lists the constants and typical values used in this study [25, 46].

TABLE I. Typical values of physical parameters

| Parameter | Quantity | Value |
|---|---|---|
| $A_{FN}$ | Fowler-Nordheim coefficient (at 4.5 eV) | $3.44 \times 10^{-7}$ A V$^{-2}$ |
| $B_{FN}$ | Fowler-Nordheim coefficient (at 4.5 eV) | $6.55 \times 10^{10}$ V m$^{-1}$ |
| $e$ | Electron charge | $1.602 \times 10^{-19}$ C |
| $m$ | Electron mass | $9.11 \times 10^{-31}$ kg |
| $\epsilon_0$ | Permittivity of vacuum | $8.854 \times 10^{-12}$ F m$^{-1}$ |
| $\hbar$ | Reduced Planck's constant | $1.05 \times 10^{-34}$ J s |
| $Q$ | Fowler Nordheim constant | $5.77 \times 10^{-29}$ J m |
| $\Phi$ | Work function | 4.5 eV |
| $E_*$ | Electric field scaling constant | $6.55 \times 10^{10}$ V m$^{-1}$ |
| $J_*$ | Current density scaling constant | $1.48 \times 10^{15}$ A m$^{-2}$ |
| $\nu_*$ | Collision frequency scaling constant | $2.54 \times 10^{15}$ s$^{-1}$ |
| $t_*$ | Time scaling constant | $3.93 \times 10^{-16}$ s |
| $\phi_*$ | Voltage scaling constant | 116.5 V |
| $x_*$ | Position scaling constant | $1.78 \times 10^{-9}$ m |
| $u_*$ | Velocity scaling constant | $4.53 \times 10^6$ m s$^{-1}$ |

Combining (3)-(5) and applying electron continuity $J = \rho u$ gives

$$\frac{d^2\bar{\phi}}{d\bar{x}^2} = \frac{\bar{J}}{\bar{u}} \tag{6}$$

and

$$\frac{d\bar{u}}{d\bar{t}} = \frac{d\bar{\phi}}{d\bar{x}} - \bar{\nu}\bar{u}. \tag{7}$$

Differentiating (7) with respect to $\bar{x}$, considering $\bar{u} = d\bar{x}/d\bar{t}$ to change variables, and combining with (6) gives

$$\bar{J} = \frac{d^2\bar{u}}{d\bar{t}^2} + \bar{\nu}\frac{d\bar{u}}{d\bar{t}}. \tag{8}$$

Solving (8) gives

$$\bar{u}(\bar{t}) = \frac{\bar{J}(\bar{\nu}\bar{t} + e^{-\bar{\nu}\bar{t}} - 1) + \bar{\nu}(\tilde{c} - \tilde{c}e^{-\bar{\nu}\bar{t}})}{\bar{\nu}^2}, \tag{9}$$

which we integrate to obtain

$$\bar{x}(\bar{t}) = \frac{\bar{J}\bar{\nu}\bar{t}(\bar{\nu}\bar{t} - 2) + 2e^{-\bar{\nu}\bar{t}}(\tilde{c}\bar{\nu} - \bar{J}) + 2(\bar{J} + \tilde{c}\bar{\nu}(\bar{\nu}\bar{t} - 1))}{2\bar{\nu}^3}, \tag{10}$$

where $\tilde{c}$ represents the electron's initial acceleration, making $\tilde{c} = \bar{E}_s$. While $\tilde{c} \neq 0$ for nonzero $u_0$ [46], $\tilde{c} = 0$ for $u_0 = 0$. This allows us to simplify the velocity at the anode $u(\bar{T}) = \bar{u}_f$ and the exact space-charge-limited $\bar{J}_{SCLC,exact}$ as

$$u_f = \frac{2\bar{D}\bar{\nu}[1 + e^{\bar{\nu}\bar{T}}(\bar{\nu}\bar{T} - 1)]}{-2 + e^{\bar{\nu}\bar{T}}(2 - 2\bar{\nu}\bar{T} + \bar{\nu}^2\bar{T}^2)} \tag{11}$$

and

$$\bar{J}_{SCLC,exact} = \frac{2\bar{D}\bar{\nu}^3 e^{\bar{\nu}\bar{T}}}{-2 + e^{\bar{\nu}\bar{T}}(2 - 2\bar{\nu}\bar{T} + \bar{\nu}^2\bar{T}^2)}, \tag{12}$$

respectively. Rearranging and integrating the force balance in (7) gives

$$\bar{V} = 2\bar{D}^2\bar{\nu}^2 e^{\bar{\nu}\bar{T}} \frac{[-6(1 + \bar{\nu}\bar{T}) + e^{\bar{\nu}\bar{T}}(6 - 3\bar{\nu}^2\bar{T}^2 + 2\bar{\nu}^3\bar{T}^3)]}{3[-2 + e^{\bar{\nu}\bar{T}}(2 - 2\bar{\nu}\bar{T} + \bar{\nu}^2\bar{T}^2)]^2}. \tag{13}$$

Equations (12) and (13) can be plotted parametrically to determine $\bar{J}_{SCLC}$ as a function of $\bar{V}$; however, this makes it difficult to obtain any intuition about the behavior between the CLL and MGL. To assess this, we appeal to the concept of vacuum capacitance. Applying continuity to (3) and rewriting in terms of the electric field $\bar{E} = -d\bar{\phi}/d\bar{x}$ yields

$$\bar{E} = \bar{J}\bar{t}. \tag{14}$$

Using capacitance, we can write [47]

$$J = \frac{Q_a}{AT} = \frac{\epsilon_0 |E(D)|}{T}, \tag{15}$$

where $Q_a$ is the total bound positive surface charge on the anode and $|E(D)|$ is the magnitude of the electric field at the anode. Nondimensionalizing (15) using (5) recovers (14) when evaluated at the anode at $\bar{t} = \bar{T}$. The electric potential for a space-charge limited gap in vacuum or in a fully collisional gap is given by

$$\bar{\phi}(\bar{x}) = \bar{V}\left(\frac{\bar{x}}{\bar{D}}\right)^{\xi}, \tag{16}$$

where $\xi = 4/3$ in vacuum [21, 47] and $\xi = 3/2$ in a fully collisional gap [48]. Thus, we conjecture [and shall show in (18)] that $4/3 \leq \xi \leq 3/2$ between vacuum and fully collisional. To determine $\xi$, we first differentiate (16) with respect to $\bar{x}$ to obtain

$$|\bar{E}| = \left|\frac{d\bar{\phi}(\bar{x})}{d\bar{x}}\right| = \frac{\xi \bar{V}}{\bar{D}}\left(\frac{\bar{x}}{\bar{D}}\right)^{\xi-1}. \tag{17}$$

Setting (14) and (17) equal, substituting (13) for $\bar{V}$, and evaluating at the anode ($\bar{x} = \bar{D}$ and $\bar{t} = \bar{T}$) yields

$$\xi = \frac{\bar{J}(\bar{T})\bar{T}\bar{D}}{\bar{V}} = \frac{3\bar{v}\bar{T}\left[-2 + e^{\bar{v}\bar{T}}(2 - 2\bar{v}\bar{T} + \bar{v}^2\bar{T}^2)\right]}{-6(1 + \bar{v}\bar{T}) + e^{\bar{v}\bar{T}}(6 - 3\bar{v}^2\bar{T}^2 + 2\bar{v}^3\bar{T}^3)}. \tag{18}$$

Note that $\xi$ is *only* a function of $\bar{v}\bar{T} = vT$, which represents the *total number of collisions in the gap* independent of any diode characteristics (e.g., voltage, current, gap distance, or collision frequency). This is important since $\xi$ provides a key metric for the transition from CL and MG. As we shall later show, $\xi$ and collisions over the diode gap play a critical role in determining the transition from the CLL to MGL. Incorporating temperature and pressure dependence into $\bar{\nu}$ would yield the identical $\bar{\nu}\bar{T}$ scaling in (16); however, incorporating energy dependence into $\bar{\nu}$ would alter the dependence on $u$ in the second term on the right-hand side of (7) and require a separate analysis for each gas. Nevertheless, the assessment for constant $\bar{\nu}$ provides an important first step in examining the implications of collisions on SCLC.

For the present assessment, for any diode size, voltage, or collision frequency, a meaningful number of collisions will cause a deviation from $J_{CL} \propto V^{3/2}$ at vacuum based on the alteration of $\xi$. Figure 1 shows that $\xi$ increases with $\bar{\nu}\bar{T}$ for $\bar{\nu}\bar{T} > 1$, approaching 4/3 as $\bar{\nu}\bar{T} \to 0$ and 3/2 as $\bar{\nu}\bar{T} \to \infty$. For example, for $\bar{\nu}\bar{T} = 2$, $\xi = 1.37$; for $\bar{\nu}\bar{T} = 10$, $\xi = 1.44$. This perturbation of $\xi$ from 4/3 with just a few collisions suggests a relatively easy perturbation from $J_{CL} \propto V^{3/2}$ upon introducing collisions. The dependence on $\nu T$ indicates that having a low collision frequency necessitates a high transit time to reach unity at vacuum. At $\bar{\nu}\bar{T} = 1$, $\xi = 1.35$, and $\xi = 1.33$ for $\bar{\nu}\bar{T} \lesssim 0.1$, indicating that $J_{CL} \propto V^{3/2}$ remains fairly accurate at low $\bar{\nu}$. Similarly, for high $\bar{\nu}$ (such as a solid or liquid, where $\bar{\nu}$ would essentially be infinite), a low $\bar{T}$, such as required by a small gap distance or high voltage, would be required to noticeably reduce $\xi$ and alter the $J_{MG} \propto V^2$ scaling.

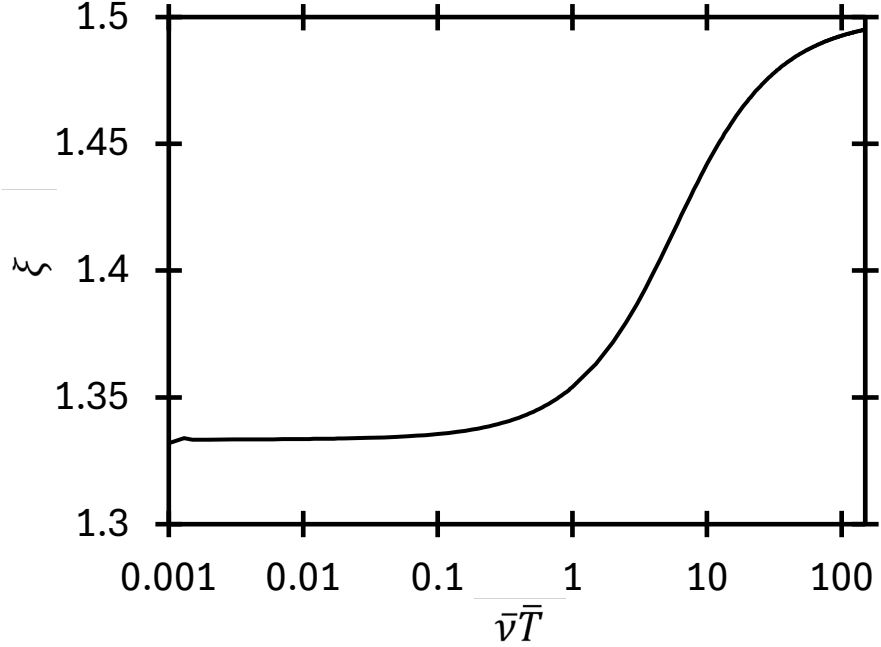

**FIG. 1**. Variation of $\xi$, which is the exponent defining the spatial variation of the electric potential $\bar{\phi}(\bar{x}) = \bar{V}(\bar{x}/\bar{D})^{\xi}$ as a function of the total number of collisions $\bar{\nu}\bar{T}$, where $\bar{\nu}$ is the nondimensional collision frequency and $\bar{T}$ is the nondimensional electron transit time.

To further probe the dependence of the general collisional SCLC on $\bar{V}$ and $\bar{D}$, we appeal to the capacitance equation from (15), which requires estimating $\bar{T}$. We first approximate the dependence of $\bar{u}$ on $\bar{x}$ by changing variables on the left-hand side of (7) to obtain

$$\bar{u}\frac{d\bar{u}}{d\bar{x}} + \bar{\nu}\bar{u} = \frac{d\bar{\phi}}{d\bar{x}}. \tag{19}$$

While this solution has analytic solutions for viscous flow under some definitions of $\bar{\phi}$ as a function of $x$ [49], the $\bar{x}^{\xi-1}$ dependence of $d\bar{\phi}/d\bar{x}$ from on the right-hand side of (19) eliminates this possibility here. To estimate the functional dependence $\bar{u}(\bar{x})$ in (19), we follow the approach in vacuum [47] by noting from (6) and continuity, given by $\bar{J} = \bar{\rho}\bar{u}$, that $d^2\bar{\phi}(\bar{x})/d\bar{x}^2 = \bar{\rho} \propto (\bar{x}/\bar{D})^{\xi-2}$. Since $\bar{J}$ is constant,

$$\bar{u}(\bar{x}) \approx \bar{u}_{max}(\bar{x}/\bar{D})^{2-\xi}. \tag{20}$$

Applying this form of $\bar{u}(\bar{x})$ to (19) yields

$$(2-\xi)\frac{\bar{u}_{max}^2}{\bar{D}}\left(\frac{\bar{x}}{\bar{D}}\right)^{3-2\xi} + \bar{v}\bar{u}_{max}\left(\frac{\bar{x}}{\bar{D}}\right)^{2-\xi} = \xi\frac{\bar{V}}{\bar{D}}\left(\frac{\bar{x}}{\bar{D}}\right)^{\xi-1}. \tag{21}$$

Solving for $\bar{u}_{max}$ at $\bar{x} = \bar{D}$, which corresponds to the maximum velocity and electric potential, gives

$$\bar{u}_{max} = \bar{D}\bar{v}\frac{-1 + \sqrt{1 + 4\xi(2-\xi)\bar{V}/(\bar{D}^2\bar{v}^2)}}{2(2-\xi)}. \tag{22}$$

As $\bar{v} \to 0$, $u_{max} \to \sqrt{\xi\bar{V}/(2-\xi)} = \sqrt{2\bar{V}}$, in agreement with the CL derivation [5]. As $\bar{v} \to \infty$, $u_{max} \to \xi\bar{V}\bar{D}^{-1}\bar{v}^{-1} = 3\bar{V}/(2\bar{D}\bar{v})$, in agreement with the expected value from MG of $u_{max} = [d\bar{\phi}(\bar{D})/d\bar{x}]/\bar{v} = 3\bar{V}/(2\bar{D}\bar{v})$ [48]. From kinematics, we can write $\bar{u} = d\bar{x}/d\bar{t}$ and solve by separation of variables, integrating the $\bar{x}$ components from $\bar{x} = 0$ to $\bar{D}$ and the $\bar{t}$ components from 0 to $\bar{T}$, to obtain

$$\bar{T} = \frac{\bar{D}}{u_{max}(\xi - 1)}. \tag{23}$$

Combining (17), (19), (22), and (23) gives

$$\bar{J}_{SCLC} = \frac{\bar{V}\bar{v}\xi(\xi-1)}{2\bar{D}(2-\xi)}\left[-1 + \sqrt{1 + \frac{4\bar{V}}{\bar{D}^2\bar{v}^2}\xi(2-\xi)}\right], \tag{24}$$

which recovers $\bar{J}_{CL} = 4\sqrt{2}\bar{V}^{3/2}/(9\bar{D}^2)$ in the limit of $\bar{v} \to 0$ and $\bar{J}_{MG} = 9\bar{V}^2/(8\bar{v}\bar{D}^3)$ in the limit of $\bar{v} \to \infty$. The second term in the radical of (24) arises from $u_{max,coll}/u_{max,vac} = \left\{[\xi\bar{V}\bar{D}^{-1}\bar{v}^{-1}]/\left[\sqrt{\xi\bar{V}/(2-\xi)}\right]\right\}^2$. When this term is small, collisions dominate; when it is large, the system behaves more like vacuum.

To quantify this modification, the binomial expansion of (24) when $\bar{V}\bar{D}^{-2}\bar{v}^{-2} \ll 1$ yields

$$\bar{J}_{SCLC,est,coll} \approx \frac{\bar{V}^2 \xi^2 (\xi - 1)}{\bar{D}^3 \bar{v}} \left[ 1 - \frac{2\bar{V}}{\bar{D}^2 \bar{v}^2} \xi (2 - \xi) + \frac{4\bar{V}^2}{\bar{D}^4 \bar{v}^4} \xi^2 (2 - \xi)^2 + O\left[ \frac{\bar{V}^4 \xi^4}{\bar{D}^4 \bar{v}^6} (2 - \xi)^4 \right] \right], \quad (25)$$

where the second and third terms in the square brackets correct $\bar{J}_{MG}$ and show the modification of the $\bar{V}^2 \bar{D}^{-3} v^{-1}$ scaling. The dependence of this modification on all the parameters suggests the complicated interplay between them when assessing the implications of collisionality. Similarly, we can quantify the deviation from $\bar{J}_{CL}$ by rewriting (24) as

$$\bar{J}_{SCLC} = \frac{\bar{V} \bar{v} \xi (\xi - 1)}{2\bar{D}(2 - \xi)} \left[ -1 + \left[ \frac{4\bar{V}}{\bar{D}^2 \bar{v}^2} \xi (2 - \xi) \right]^{1/2} \sqrt{1 + \frac{\bar{D}^2 \bar{v}^2}{4\bar{V} \xi (2 - \xi)}} \right], \quad (26)$$

which we expand in terms of $\bar{V}^{-1} \bar{D}^2 \bar{v}^2 \ll 1$ to obtain

$$\bar{J}_{SCLC,vac} = \frac{\bar{V}^{3/2} \xi^{3/2} (\xi - 1)}{\bar{D}^2 (2 - \xi)^{1/2}} \left[ 1 + \frac{\bar{D}^2 \bar{v}^2}{8\bar{V} \xi (2 - \xi)} - \frac{\bar{D}^4 \bar{v}^4}{128 \bar{V}^2 \xi^2 (2 - \xi)^2} + O\left[ \frac{\bar{D}^6 \bar{v}^6}{128 \bar{V}^4 \xi^4 (2 - \xi)^4} \right] \right]. \quad (27)$$

To determine $\bar{J}_{SCLC}$ using (24), we fix $\bar{D}, \bar{v}$, and $\bar{V}$. Substituting (22) into (23) gives $\bar{T}$ as a function of $\xi$ as

$$\bar{T} = \frac{2(2 - \xi)}{\bar{v}(\xi - 1)} \frac{1}{-1 + \sqrt{1 + 4\xi(2 - \xi)\bar{V}/(\bar{D}^2 \bar{v}^2)}}, \quad (28)$$

which we substitute into (18) to obtain $\xi$ numerically. We then substitute $\xi$ into (28) to obtain $\bar{T}$ and (24) to obtain $\bar{J}_{SCLC}$.

Figure 2 assesses the estimated $\bar{J}_{SCLC}$ as a function of $\bar{V}$ for $\bar{D} = 1000$ and $\bar{v} = 10/7$ to compare to the exact solution [46]. Figure 2a compares the capacitance calculation from (24) with the low and high voltage asymptotes from (25) and (27). The asymptotes agree with (24) in the appropriate voltage regimes. Defining $\chi = \bar{J}_{SCLC}/\bar{J}_{SCLC,exact}$ to compare the SCLC obtained using capacitance ($\bar{J}_{SCLC}$) to the exact SCLC $\bar{J}_{SCLC,exact}$ using (12) and (13) [46] for a given $\bar{V}$ shows that the two calculations agree within 6%. Figure 2b also shows that $\xi$ varies from 1.5 at low voltages, corresponding to the MGL, to ~1.33 at high voltages, corresponding to the CLL.

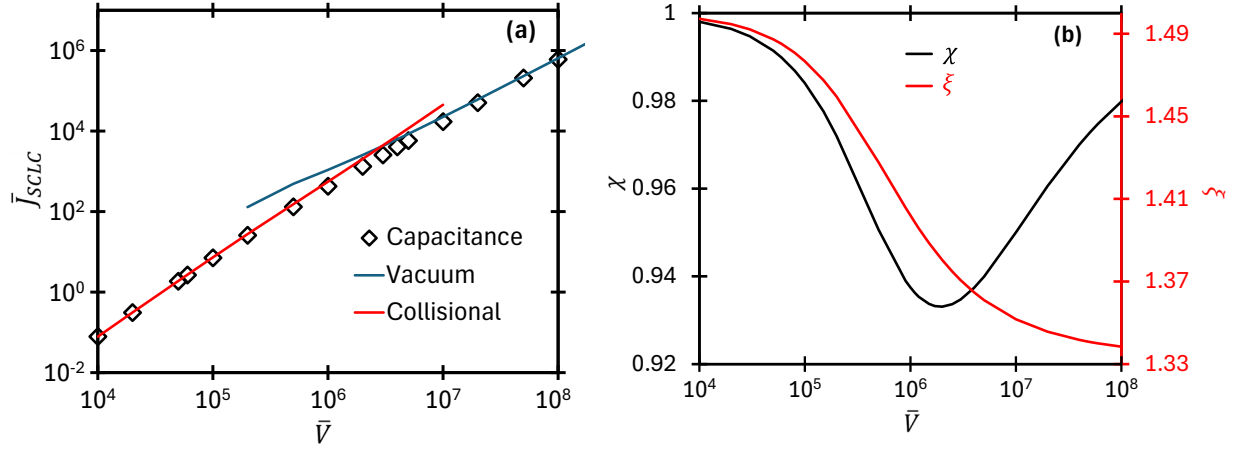

**FIG. 2.** (a) SCLC $\bar{J}_{SCLC}$ as a function of applied voltage $\bar{V}$ calculated using capacitance from (24) compared to the collisional (25) and vacuum (27) asymptotes for $\bar{D} = 1000$ and $\bar{v} = 10/7$. (b) Comparison of $\bar{J}_{SCLC}$ determined using vacuum capacitance to the exact solution $\bar{J}_{SCLC,exact}$ by $\chi = \bar{J}_{SCLC}/\bar{J}_{SCLC,exact}$ and the electric potential spatial exponent $\xi$ as a function of $\bar{V}$.

Figure 3 assesses the estimated $\bar{J}_{SCLC}$ as a function of $\bar{V}$ for $\bar{D} = 1000$ and $\bar{v} = 1/70$ to compare to the exact results in a near-vacuum environment [46]. Figure 3a shows the agreement of the estimate from the capacitance-based calculations to the respective asymptotic solutions for $\bar{V} \ll 1$ and $\bar{V} \gg 1$. Compared to the higher $\bar{v}$ case from Fig. 2a, $\bar{V}$ at the transition from the MG-like behavior to CL-like behavior occurs at a lower $\bar{V}$ in Fig. 3a because the lower collisionality makes the initial behavior more closely resemble vacuum. Figure 3b shows similar agreement between $\bar{J}_{SCLC}$ and $\bar{J}_{SCLC,exact}$ over the voltage range studied. The asymptotic behavior of $\xi$ also resembles Fig. 2b with voltage, although numerical stability at the low $\bar{v}$ made it difficult for us to use a sufficiently low $\bar{V}$ to achieve $\xi = 1.50$.

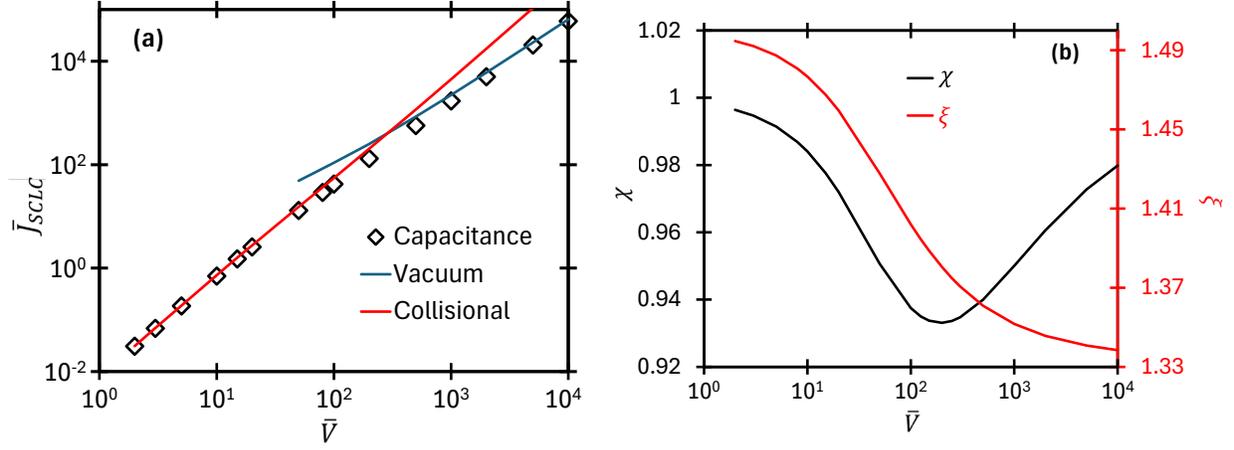

**FIG. 3.** (a) SCLC $\bar{J}_{SCLC}$ as a function of applied voltage $\bar{V}$ calculated using capacitance from (24) compared to the collisional (25) and vacuum (27) asymptotes for $\bar{D} = 1000$ and $\bar{v} = 1/70$. (b) Comparison of $\bar{J}_{SCLC}$ determined using vacuum capacitance to the exact solution $\bar{J}_{SCLC,exact}$ by $\chi = \bar{J}_{SCLC}/\bar{J}_{SCLC,exact}$ and the electric potential spatial exponent $\xi$ as a function of $\bar{V}$.

We next assess the transition between the collisional and vacuum asymptotes by matching the first order terms from (25) and (27) to obtain the transition voltage as

$$\bar{V}_{trans} = \frac{(\bar{D}\bar{v})^2}{(2 - \xi_{trans})\xi_{trans}}, \qquad (29)$$

where $\xi_{trans}$ corresponds to $\xi$ at the transition between the two asymptotes. This nexus relationship comes physically from the ratio of the maximum velocity in a vacuum space-charge limited gap to a fully collisional space-charge limited gap. Combining (18), (22), (23), and (29) gives $\xi_{trans} = 1.40$. This indicates that $\xi_{trans}$ is independent of the diode ($\bar{V}$ and $\bar{D}$) and gas ($\bar{v}$) characteristics. Figure 4 shows (29) and the voltage scaling with $(\bar{D}\bar{v})^2$.

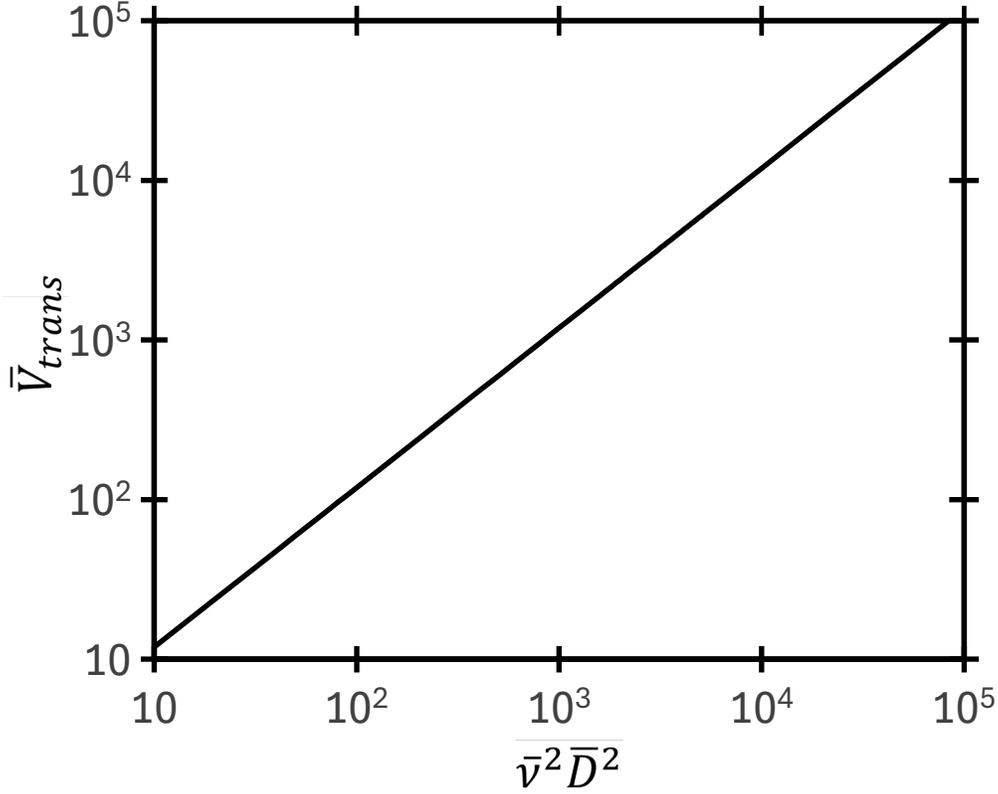

**FIG. 4.** Voltage $\bar{V}_{trans}$ corresponding to the transition from the first order term of (25) to the first order term of (27) as a function of $\bar{\nu}^2 \bar{D}^2$, where $\bar{\nu}$ is dimensionless collision frequency and $\bar{D}$ is dimensionless gap distance.

This approach may also be extended to multiple dimensions following the process demonstrated previously in vacuum [21]. Assuming that (24) is given in the $w$-plane with the normalized locations of the anode and the cathode at $u = \bar{u}_A$ and $u = \bar{u}_C$, respectively, gives the SCLC with $\bar{D} = \bar{u}_A - \bar{u}_C$ as

$$\bar{J}_{SCLC} = \frac{\bar{V}\bar{\nu}\xi(\xi-1)}{2(\bar{u}_A - \bar{u}_C)(2-\xi)}\left[-1 + \sqrt{1 + \frac{4\bar{V}}{(\bar{u}_A - \bar{u}_C)^2\bar{\nu}^2}\xi(2-\xi)}\right]. \tag{30}$$

If a given geometry in the $z$-plane can be mapped conformally onto a planar 1D geometry using the mapping function $w = f(z)$ in the $w$-plane, we may use (30) to calculate the SCLC in the $z$- plane. We can map a 3D diode with anode and cathode of width $W$ and length $L$ separated by a distance $D$ conformally onto the superposition of two 1D diodes using Schwarz-Christoffel transformations to obtain the capacitance [50]. Transforming (30) to the $w$-plane yields

$$\bar{J}_{SCLC} = \frac{\bar{V}\bar{v}\xi(\xi-1)}{2\mathcal{D}(2-\xi)}\left[-1 + \sqrt{1 + \frac{4\bar{V}}{\mathcal{D}^2\bar{v}^2}\xi(2-\xi)}\right], \quad (31)$$

where $\mathcal{D}$ is the modified gap distance because of the fringing fields and is given by

$$\mathcal{D} = \left[\frac{F(\pi/2,\mu_1)}{WDF(\pi/2,m_1)} + \frac{F(\pi/2,\mu_2)}{LDF(\pi/2,m_2)} - \frac{1}{D^2}\right]^{-1/2}, \quad (32)$$

where $F(\pi/2,\mu)$ and $E(\pi/2,\mu)$ are the complete elliptic integrals of the first and second kind, respectively, with $F(\theta,\mu) = \int_0^\theta dt/\sqrt{1-\mu^2\sin^2 t}$ and $E(\theta,\mu) = \int_0^\theta \sqrt{1-\mu^2\sin^2 t}\,dt$ as the incomplete elliptic integrals of the first and second kind, respectively [21]. Assuming two moduli $m_1$ and $m_2$ such that $m_1, m_2 \in (0,1)$ gives the corresponding complementary moduli $\mu_1$ and $\mu_2$ from $\mu_i^2 \equiv 1 - m_i^2$ for $i = \{1,2\}$. We then construct the arguments $\omega_1$ and $\omega_2$ using

$$\sin^2 \omega_i = \frac{F\left(\frac{\pi}{2},\mu_i\right) - E\left(\frac{\pi}{2},\mu_i\right)}{\mu_i^2 F\left(\frac{\pi}{2},\mu_i\right)}. \quad (33)$$

Using the pairs of $(m_1,\mu_1)$ and $(m_2,\mu_2)$ yields

$$\frac{W}{D} = \frac{F\left(\frac{\pi}{2},\mu_1\right)E(\omega_1,\mu_1) - E\left(\frac{\pi}{2},\mu_1\right)F(\omega_1,\mu_1)}{\left[E\left(\frac{\pi}{2},\mu_1\right) - F\left(\frac{\pi}{2},\mu_1\right)\right]F\left(\frac{\pi}{2},m_1\right) + F\left(\frac{\pi}{2},\mu_1\right)E\left(\frac{\pi}{2},m_1\right)} \quad (34)$$

and

$$\frac{L}{D} = \frac{F\left(\frac{\pi}{2},\mu_2\right)E(\omega_2,\mu_2) - E\left(\frac{\pi}{2},\mu_2\right)F(\omega_2,\mu_2)}{\left[E\left(\frac{\pi}{2},\mu_2\right) - F\left(\frac{\pi}{2},\mu_2\right)\right]F\left(\frac{\pi}{2},m_2\right) + F\left(\frac{\pi}{2},\mu_2\right)E\left(\frac{\pi}{2},m_2\right)}. \quad (35)$$

When $W \gg D$ and $L \gg D$, as is often the case [36-41], the equivalent distance from (32) is given by [21]

$$\mathcal{D} \approx \left[ \frac{1}{D^2} + \frac{1 + \ln\left(\frac{2\pi W}{D}\right)}{\pi WD} + \frac{1 + \ln\left(\frac{2\pi L}{D}\right)}{\pi LD} \right]^{-1/2}, \tag{36}$$

which reduces to $\mathcal{D} \approx D$ when $W \to \infty$ and $L \to \infty$, as expected. Thus, if the general collisional SCLC is a better fit to the experimentally obtained $J - V$ curves, the mobility can be more accurately estimated from (31) with an equivalent gap distance.

In summary, this Letter predicts SCLC from CL to MG by using vacuum capacitance in good agreement with the previously derived exact solution. Our derivation shows that the space-charge limited potential *always* exhibits $\bar{\phi}(\bar{x}) \propto (\bar{x}/\overline{D})^{\xi}$ scaling with $\xi$ ranging from 4/3 for vacuum to 3/2 for a fully collisional gap. This scaling provides a means to assess the collision frequency (or electron mobility) in a space-charge limited diode. For instance, particle-in-cell (PIC) simulations with gas pressure as an input and $\phi(x)$ as an output can be fit to $(x/D)^{\xi}$ to obtain $\xi$ to characterize the collisonality. Substituting $\xi$ into (18) gives the number of collisions $\nu T$ and, by using $T$ from PIC simulations, the collision frequency $\nu$. The nexus between the asymptotic solutions for vacuum and collisional diodes depends on matching the maximum velocity of vacuum and collisional conditions, giving $\bar{V} \propto \bar{\nu}^2 \overline{D}^2$ at the nexus. The universality of $\xi = 1.40$ at this transition indicates that nexus always occurs for the *same number of collisions* in the gap, although this behavior is more likely to be considered as a probability-based property (such as mean free path) than as a deterministic one, suggesting that $\bar{\nu}\bar{T} < 1$ may still have meaning as a statistical condition of minimal collisions. For semiconductors, using the MGL to assess electron mobility assumes a trap-free diode [51]; however, incorporating traps leads to the Mark-Helfrich law (MHL) [52]. Incorporating this collisional dependence may further elucidate studies of the MGL and MHL using capacitance [48] and nexus theory [53], as well as for systems comprised of both solids and vacuum [54].

This work was supported in part by SCALE: U.S. Department of Defense Contract No. W52P1J-22-9-3009, the Joint Directed Energy Transition Office under contract number HQ0642384797, and Sandia National Laboratories (SNL) under contract number PO2516032. SNL is a multi-mission laboratory managed and operated by National Technology & Engineering Solutions of Sandia, LLC (NTESS), a wholly owned subsidiary of Honeywell International Inc., for the U.S. Department of Energy's National Nuclear Security Administration (DOE/NNSA) under contract DE-NA0003525.

**AUTHOR DECLARATIONS**

**Conflict of Interest**

The authors have no conflicts to disclose.

**Author Contributions**

**Allen L. Garner:** Conceptualization (Lead); Formal Analysis (Lead); Funding acquisition (Lead); Investigation (Lead); Methodology (Lead); Project Administration (Lead); Supervision (Lead); Writing – original draft (lead); Writing – review & editing (Equal). **N. R. Sree Harsha:** Conceptualization (Supporting); Investigation (Supporting); Methodology (Supporting); Writing – original draft (Supporting); Writing – review & editing (Equal).

**DATA AVAILABILITY**

The data that support the findings of this study are available from the corresponding author upon reasonable request.

**REFERENCES**

1. P. Zhang, Y. S. Ang, A. L. Garner, Á. Valfells, J. W. Luginsland, and L. K. Ang, Space-charge limited current in nanodiodes: Ballistic, collisional and dynamical effects. J. Appl. Phys. **129**, 100902 (2021).

2. K. L. Jensen, A tutorial on electron sources. IEEE Trans. Plasma Sci. **46**, 1881–1899 (2018).

3. A. L. Garner, G. Meng, Y. Fu, A. M. Loveless, R. S. Brayfield II, and A. M. Darr, Transitions between electron emission and gas breakdown mechanisms across length and pressure scales. J. Appl. Phys. **128**, 210903 (2020).

4. A. L. Garner, A. M. Loveless, A. M. Darr, and H. Wang, "Gas Discharge and Electron Emission for Microscale and Smaller Gaps," in *Pulsed Discharge Plasmas: Characteristics and Applications,* edited by T. Shao and C. Zhang, (Springer, Singapore Pte Ltd., 2023), Chap. 3, pp. 75-95.

5. Y. Y. Lau, Y. Liu, and R. K. Parker, Electron emission: From the Fowler-Nordheim relation to the Child-Langmuir law. Phys. Plasmas **1**, 2082-2085 (1994).

6. A. M. Darr, A. M. Loveless, and A. L. Garner, Unification of field emission and space charge limited emission with collisions. Appl. Phys. Lett. **114**, 014103 (2019).

7. A. M. Darr, C. R. Darr, and A. L. Garner, Theoretical assessment of transitions across thermionic, field, and space-charge limited emission. Phys. Rev. Res. **2**, 033137 (2020).

8. L. I. Breen, A. M. Loveless, A. M. Darr, K. L. Cartwright, and A. L. Garner, The transition from field emission to collisional space-charge limited current with nonzero initial velocity. Sci. Rep. **13**, 14505 (2023).

9. K. L. Jensen, M. McDonald, O. Chubenko, J. R. Harris, D. A. Shiffler, N. A. Moody, J. J. Petillo, and A. J. Jensen, Thermal-field and photoemission from meso- and micro-scale features: Effects of screening and roughness on characterization and simulation. J. Appl. Phys. **125**, 234303 (2019).

10. S. A. Lang, A. M. Darr, and A. L. Garner, Incorporating photoemission into the theoretical unification of electron emission and space-charge limited current. J. Vac. Sci. Technol. B **39**, 062808 (2021).

11. P. Zhang, A. Valfells, L. K. Ang, J. W. Luginsland, and Y. Y., 100 years of the physics of diodes. Appl. Phys. Rev. **4**, 011304 (2017).

12. C. D. Child, Discharge from hot CaO. Phys. Rev. Ser. I **32**, 492–511 (1911).

13. I. Langmuir, The effect of space charge and residual gases on thermionic currents in high vacuum. Phys. Rev. **2**, 450–486 (1913).

14. A. L. Garner, A. M. Darr, and N. R. Sree Harsha, A tutorial on calculating space-charge limited current density for general geometries and multiple dimensions. IEEE Trans. Plasma Sci. **50**, 2528-2540 (2022).

15. A. M. Darr and A. L. Garner, A coordinate system invariant formulation for space-charge limited current in vacuum. Appl. Phys. Lett. **115**, 054101 (2019).

16. N. R. Sree Harsha and A. L. Garner, Applying conformal mapping to derive analytical solutions of space-charge-limited current density for various geometries. IEEE Trans. Electron Devices **68**, 264-270 (2021).

17. N. R. Sree Harsha, J. M. Halpern, A. M. Darr, and A. L. Garner, Space-charge-limited current density for nonplanar diodes with monoenergetic emission using Lie-point symmetries. Phys. Rev. E **106**, L063201 (2022).

18. N. R. Sree Harsha and A. L. Garner, Analytic solutions for space-charge-limited current density from a sharp tip. IEEE Trans. Electron Devices **68**, 6525-6531 (2021).

19. J.W. Luginsland, Y.Y. Lau, and R.M. Gilgenbach, Two-dimensional Child-Langmuir law. Phys Rev Lett **77**, 4668–4670 (1996).

20. Y.Y. Lau, Simple theory for the two-dimensional Child-Langmuir law. Phys. Rev. Lett. **87(27)**, 278301 (2001).

21. N. R. Sree Harsha, M. Pearlman, J. Browning, and A. L. Garner, A multi-dimensional Child-Langmuir law for any diode geometry. Phys. Plasmas **28**, 122103 (2021).

22. I. Langmuir, The effect of space charge and initial velocities on the potential distribution and thermionic current between parallel plane electrodes. Phys. Rev. **21**, 419 (1923).

23. G. Jaffé, On the currents carried by electrons of uniform initial velocity. Phys. Rev. **65**, 91–98 (1944).

24. S. Liu and R. A. Dougal, Initial velocity effect on space-charge-limited currents. J. Appl. Phys. **78**, 5919–5925 (1995).

25. P. V. Akimov, H. Schamel, H. Kolinsky, A. Y. Ender, and V. I. Kuznetsov, The true nature of space-charge-limited currents in electron vacuum diodes: A Lagrangian revision with corrections. Phys. Plasmas **8**, 3788-3798 (2001).

26. J. M. Halpern, A. M. Darr, N. R. S. Harsha, and A. L. Garner, A coordinate system invariant formulation for space-charge limited current with nonzero injection velocity. Plasma Sources Sci. Technol. **31**, 095002 (2022).

27. T. Lafleur, Space-charge limited current with a finite injection velocity revisited. Plasma Sources Sci. Technol. **29**, 065002 (2020).

28. J. B. Huang, R. H. Yao, P. Zhao, and Y. B. Zhu, Simulation of space-charge-limited current for hot electrons with initial velocity in a vacuum diode. IEEE Trans. Electron Devices **68**, 3604-3610 (2021).

29. K. H. Schoenbach and K. Becker, 20 years of microplasma research: a status report. Eur. Phys. J. **70**, 29 (2016).

30. Y. Fu, P. Zhang, J. P. Verboncoeur, and X. Wang, Electrical breakdown from macro to micro/nano scales: a tutorial and a review of the state of the art. Plasma Res. Expr. 2, 013001 (2020).

31. D. B. Go and A. Venkattraman, Microscale gas breakdown: ion-enhanced field emission and the modified Paschen's curve. J. Phys. D: Appl. Phys. **47**, 503001 (2014).

32. A. L. Garner, A. M. Loveless, J. N. Dahal, and A. Venkattraman, A tutorial on theoretical and computational techniques for gas breakdown in microscale gaps. IEEE Trans. Plasma Sci. **48**, 808-824 (2020).

33. N. F. Mott and R. W. Gurney, *Electronic Processes in Ionic Crystals* (Clarendon, Oxford, 1940).

34. H. Wang, R. S. Brayfield II, A. M. Loveless, A. M. Darr, and A. L. Garner, Experimental study of gas breakdown and electron emission in nanoscale gaps at atmospheric pressure. Appl. Phys. Lett. **120**, 124103 (2022).

35. A. M. Loveless, A. M. Darr, and A. L. Garner, Linkage of electron emission and breakdown mechanism theories from quantum scales to Paschen's law. Phys. Plasmas **28**, 042110 (2021).

36. V. M. Le Corre, E. A. Duijnstee, O. El Tambouli, J. M. Ball, H. J. Snaith, J. Lim, and L. J. A. Koster, Revealing charge carrier mobility and defect densities in metal halide perovskites via space-charge-limited current measurements. ACS Energy Lett. **6**, 1087-1094 (2021).

37. J. M. Montero, J. Bisquert, G. Garcia-Belmonte, E. M. Barea, and H. J. Bolink, Trap-limited mobility in space-charge limited current in organic layers. Org. Electron. **10**, 305–312 (2009).

38. J. Dacuna and A. Salleo, Modeling space-charge-limited currents in organic semiconductors: Extracting trap density and mobility. Phys. Rev. B **84**, 195209 (2011).

39. V.D. Mihailetchi, J. Wildeman, and P. W. M. Blom, Space-charge limited photocurrent. Phys. Rev. Lett. **94**, 126602 (2005).

40. D. S. Chung, D. H. Lee, C. Yang, K. Hong, C. E. Park, J. W. Park, and S-K. Kwon, Origin of high mobility within an amorphous polymeric semiconductor: Space-charge-limited current and trap distribution. Appl. Phys. Lett. **93**, 033303 (2008).

41. M. S. Alvar, P. W. M. Blom, and G. J. A. H. Wetzelaer, Space-charge-limited electron and hole currents in hybrid organic-inorganic perovskites. Nat. Commun. **11**, 4023 (2020).

42. M. S. Benilov, The Child-Langmuir law and analytical theory of collisionless to collision-dominated sheaths. Plasma Sources Sci. Technol. **18**, 014005 (2008).

43. M. S. Benilov, Collision-dominated to collisionless electron-free space-charge sheath in a plasma with variable ion temperature. Phys. Plasmas **7**, 4403-4411 (2000).

44. D. Chernin, Y. Y. Lau, J. J. Petillo, S. Ovtchinnikov, D. Chen, A. Jassem, R. Jacobs, D. Morgan, and J. H. Booske, Effect of nonuniform emission on Miram curves. IEEE Trans. Plasma Sci. **48**, 146-155 (2020).

45. D. Chen, R. Jacobs, D. Morgan, and J. Booske, Physical factors governing the shape of the Miram curve knee in thermionic emission. IEEE Trans. Electron Devices **70**, 1219-1225 (2023).

46. L. I. Breen and A. L. Garner, Collisional space-charge limited current with monoenergetic velocity: From Child-Langmuir to Mott-Gurney. Phys. Plasmas **31**, 032102 (2024).

47. R. J. Umstattd, C. G. Carr, C. L. Frenzen, J. W. Luginsland, and Y. Y. Lau, A simple physical derivation of Child–Langmuir space-charge-limited emission using vacuum capacitance. Am. J. Phys. **73**, 160–163 (2005).

48. Y. B. Zhu and L. K, Ang, Analytical re-derivation of space charge limited current in solids using capacitor model. J. Appl. Phys. **110**, 094514 (2011).

49. A. Ali, D. N. K. Marwat, and S. Asghar, New approach to the exact solution of viscous flow due to stretching (shrinking) and porous sheet. Results Phys. **7**, 1122-1127 (2017).

50. R. E. Matick and A. E. Ruehli, Accurate 3-D capacitance of parallel plates from 2-d analytical superposition. IEEE Trans. Compon. Packag. Manuf. Technol. **3**, 299-305 (2013)

51. J. A. Rohr, D. Moia, S. A. Haque, T. Kirchartz, and J. Nelson, Exploring the validity and limitations of the Mott–Gurney law for charge-carrier mobility determination of semiconducting thin-films. J. Phys.: Condens. Matt. **30**, 105901 (2018).

52. P. Mark and W. Helfrich, Space-charge-limited currents in organic crystals. J. Appl. Phys. **33**, 205-215 (1962).

53. C. Chua, Y. S. Ang, and L. K. Ang, Tunneling injection to trap-limited space-charge conduction for metal-insulator injunction. Appl. Phys. Lett. **121**, 192109 (2022).

54. Y. B. Zhu, K. Geng, Z. S. Cheng, and R. H. Yao, Space-charge-limited current injection into free space and trap-filled solid. IEEE Trans. Plasma Sci. **49**, 2107-2112 (2021).